\documentclass[twocolumn,showpacs,preprintnumbers,amsmath,amssymb,prb]{revtex4}

\usepackage{graphicx}
\usepackage[usenames]{color}
\usepackage{dcolumn}
\usepackage{bm}
\usepackage{color}

\begin{document}

\title{One-Dimensional Quantum Channel in Graphene Line Defect}
\author{Juntao Song$^{1}$, Haiwen Liu$^{2}$, Hua Jiang$^{3}$, Qing-feng Sun$^{2}$  and X. C. Xie$^{3}$}
\affiliation{$^1$Department of Physics and Hebei Advanced Thin Film  Laboratory, Hebei Normal University, Hebei
050024, China \\
$^2$Beijing National Lab for Condensed Matter Physics and Institute
of Physics, Chinese Academy of Sciences, Beijing 100080, China \\
$^3$International Center for Quantum Materials, Peking University, Beijing
10087, China}

\date{\today}

\begin{abstract}
Using a tight binding model, we study a line defect in graphene
where a bulk energy gap is opened by sublattice symmetry breaking.
It is found that sublattice symmetry breaking may induce many
configurations that correspond to different band spectra. In
particular, a gapless state is observed for a configuration which
hold a mirror symmetry with respect to the line defect. We find that
this gapless state originates from the line defect and is
independent of the width of the graphene ribbon, the location of the
line defect, and the potentials in the edges of ribbon. In particular,
the gapless state can be controlled by the gate voltage
embedded below the line defect. Finally, this result is supported
with conductance calculations. This study shows how a quantum
channel could be constructed using a line defect and how the quantum channel
can be controlled by tuning the gate voltage embedded below the line
defect.
\end{abstract}


\maketitle

\section{Introduction}
Since the experimental discovery of
graphene,~\cite{Novoselov0,Novoselov,Geim1,Geim2} extensive
attention~\cite{Avouris,Neto0,Bolotin,Balandin,Abergel,Schwierz} has
been given to this field due to graphene's unique band structure and
properties. It has been shown in many studies that pristine graphene
is a zero-gap semiconductor and has a linear dispersion
relationship nearby the Dirac points,~\cite{Avouris,Neto0} which makes electrons behave as
relativistic Dirac particles. Moreover, it has been experimentally
verified that graphene has a remarkably high electron mobility at
room temperature, with reported values in excess of
$15,000cm^2V^{-1}s^{-1}$.~\cite{Geim1} High electron mobility makes
graphene an excellent conductor. Owing to its exceptional
electrical~\cite{Bolotin} and thermal~\cite{Balandin} transport
properties, graphene has been an important two-dimensional material
for exploring condensed matter physical
phenomena~\cite{Geim1,Neto0,Abergel} and is expected to be very
useful in the next generation of electronic devices.~\cite{Schwierz}

Before applications of graphene can be realized in electronic
devices, control over the electron transport in graphene needs to be
improved. Inspired by the role of defects, vacancies, and dopants in
the semiconductor industry, many researchers have studied the
effects related to disorder in graphene, which could be due to
adsorbed atoms (or molecules), charged impurities, vacancies or
other topological defects.~\cite{Neto0,Peres0}

Point defects~\cite{Lusk1,Lusk2,Carr,Banhart}, and magnetic behavior
near point
defects,~\cite{Esquinazi,Lehtinen,Vozmediano,Okada0,Yazyev0,Wang} in
graphene have been extensively studied in many papers. Given
carbon's lack of $d$ or $f$ electrons, magnetism in graphene seems
unlikely. Magnetic behavior
near point defects has been of increasing
interest, and has attracted extensive
attention.~\cite{Cervenka,Ugeda,Chang,Chen} In addition, because of
universality of grain boundary in graphene, it has been investigated
and
is predicted to have distinct electronic,~\cite{Cervenka2,Peres,Yazyev1,Mesaros}
magnetic,~\cite{Cervenka} chemical~\cite{Malola} and
mechanical~\cite{Liu,Grantab,Yazyev2} properties that strongly
depend on the atomic arrangement. Furthermore, it has also been
found that adsorbed atoms (or molecules) greatly affect the
properties of graphene, such as lithium, aluminum, iron, and gold
adatoms in graphene.~\cite{Weeks, HBZhang,Chan,Ding} The previous
researches in this subject illustrate that the properties of
graphene can be controlled by defects, grain boundaries, or adsorbed
atoms.

Recently, a peculiar topological line defect in graphene was
reported experimentally by Lahiri et al.~\cite{Lahiri} This
topological line defect is created by alternating the
Stone-Thrower-Wales defect~\cite{Thrower} and divacancies, leading
to a pattern of repeating paired pentagons and octagons, as shown in
Fig. 1. Importantly, it is found in this experiment that this line
defect has metallic characteristics. Subsequently, Gunlycke and
White~\cite{Gunlycke} proposed a valley filter based on scattering
off this line defect in graphene. Using a tight-binding model
calculation, metallic characteristics and Fabry-Perot oscillation
phenomena have been observed in graphene line defects by Bahamon et
al.~\cite{Bahamon} By using first-principles calculations,
researchers from two different groups have calculated the electronic
structure of graphene with topological line defects, and predicted a
possible ferromagnetic ordering in line defects.~\cite{Okada,Kou} In
addition, a zigzag graphene nanoribbon edge reconstruction with
Stone-Wales defects has also been studied recently by
Rodrigues et al.~\cite{Rodrigues}

Note that this line
defect manifests a metallic behavior in most of reported results.
However, because the states of line defect and the bulk states are mixed together in energy,
it is difficult to distinguish each other in experiment.
In recent, it is reported by a lot of research groups that a energy gap can be generated by various method,
e.g. adatoms or substrate. These studies inspire us that the line defect in graphene can be
observed directly if the state of line defect falls into the energy gap,
and even some promising electronic devices can be fabricated.
Motivated by the idea of finding a potential application of line defect, we therefore extend the investigation of line
defects in graphene with sublattice symmetry breaking potentials. This paper describes how some novel states are induced in the bulk gap by line
defects and that these states can be tuned continuously by changing
the gate voltage embedded below the line defect. Based on these results, a quantum channel is proposed at the end of this paper.

The rest of this paper is organized as follows. In Sec. II, the
tight-binding Hamiltonian of graphene with a line defect is
introduced. Section III
gives numerical results along with some discussions. Finally, a
brief summary is presented in Sec. IV.

\section{THEORETICAL MODEL AND FORMULA}
In the tight binding approximation with nearest neighbor
hopping energy $t$, a single layer of graphene can be described by the following Hamiltonian:
\begin{eqnarray}
H & =&-t\sum_{\langle\mathbf{i},\mathbf{j}\rangle}(c^+_\mathbf{i}c_\mathbf{j}+c^+_\mathbf{j}c_\mathbf{i})
+\sum_{\mathbf{i}}U_\mathbf{i}c^+_\mathbf{i}c_\mathbf{i},
\end{eqnarray}
where $c_\mathbf{i}$ ($c^\dagger_\mathbf{i}$) is the electron
annihilation (creation) operator on the site $\mathbf{i}$, and
$\sum_{\langle\mathbf{i},\mathbf{j}\rangle}$ sums only over the nearest neighbor
sites.

For this Hamiltonian, we consider a general situation where the bulk
lattice can be subject to a staggered sublattice potential: $U_\mathbf{i}=\Delta/2$ for lattice sites ($\circ$), $U_\mathbf{i}=-\Delta/2$ for lattice ($\bullet$), and $U_\mathbf{i}=U_d$ for the line defect sites (red), sketched in Fig.1. For simplicity, in the following description of this paper, the sublattice A is introduced to represent the lattice sites ($\circ$); the sublattice B is used to represent the lattice sites ($\bullet$). Note that A (or B) sublattice sites can possess either positive or negative staggered potentials by changing the sign of the parameter $\Delta$. Due to rearranging of carbon atoms near
the line defect, the bond distances rearrange from 1.38 to 1.44
${\AA}$, which therefore induces a variation less than $5\%$ in the
hopping term $t$. As described in almost all reported
results,~\cite{Bahamon,Okada,Kou} the hopping term $t$ can be
considered nearly unaffected with respect to a defect-free graphene
(for which $t \thickapprox 2.7 eV$). To test the results from our
model, we calculate the energy band for a nanoribbon with the
extended defect shown in Fig. 2. The calculated results arising from
the band in Fig. 2(a) show a good match with results in another
paper reported by Bahamon et al.~\cite{Bahamon}

It is worth noting that a staggered potential has had extensive
theoretical investigation in many systems. In 1991,
Haldane~\cite{Haldane} introduced this potential in a hexagonal honeycomb
lattice and proposed the existence of quantum Hall effect in the
absence of an external magnetic field. Following the Haldane model
above, Kane and Mele~\cite{Kane} studied a graphene model with a
staggered potential between different sublattices and showed
that it could be possible to observe quantum spin Hall effect due to
the intrinsic spin-orbital coupling in graphene. Although a
staggered potential can not give rise to topological quantum states,
it makes easier to observe them and is helpful to find their physical origins.
In addition, it should be noted that a quantum phase
transition can be induced by tuning the staggered potential. Therefore,
in theory, a staggered potential is significant in finding
some fascinating quantum phenomenon and quantum phase transitions,
and has been also discussed in other systems.~\cite{Yao,Son}

In experiment, staggered potentials at different sublattice sites
could be introduced in graphene by an asymmetric interaction with a
substrate. For example, when a graphene sheet is put on top of a
lattice-matched hexagonal boron nitride substrate, a staggered
potential can be observed due to the inequivalence of the two carbon
sites, one of which is on top of a boron atom and the other centers
above a boron-nitride ring.~\cite{Giovannetti} The resulting
staggered potential is also natural for silicene~\cite{CCLiu} because
silicene has a two-dimensional low-buckled honeycomb structure, which leads to
a definitely asymmetric interactions between the different
sublattice sites and the substrate. For the system of graphene with
a line defect, a staggered potential may exist due to the two
different adsorption geometries, namely the fcc and hcp, when
stacking graphene on top of Ni(111).~\cite{Lahiri} Besides,
different adatoms or deformations on different sublattices of
graphene can also lead to a staggered potential experimentally. In
brief, due to a promising realization of a staggered potential
experimentally, the staggered potential for graphene considered in
our model is reasonable.

\begin{figure}[!ht]
\includegraphics[width=1.0\columnwidth, viewport=0 0 490 570, clip]{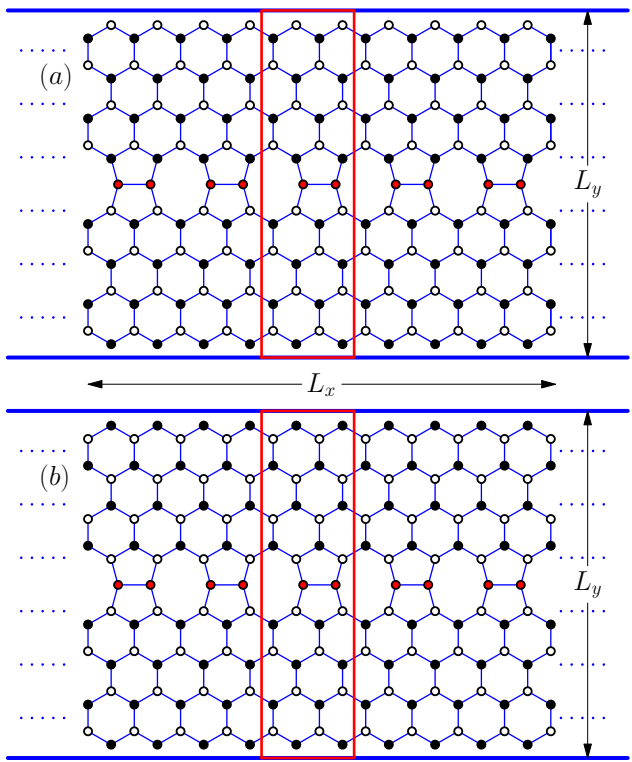}
\caption{ (Color online) A schematic diagram for an infinitely long
zigzag graphene ribbon with a line defect. There are two cases; one
is that the line defect connects the A $(\circ)$ and B $(\bullet)$
sublattice sites (a), and the other is that the line defect connects
two of the same type of sublattice sites (A-A) (b).}
\end{figure}

\section{Numerical Results and Discussion}
From the above tight-binding model for the situation shown in Fig.
1, it is easy to get a ribbon geometry with zigzag edges.
The width of the ribbon is denoted by integer M. For example, in Fig. 1,
the width $L_y \approx M\times 3a$ with $M=4$. Here, $a$ represents
carbon-carbon distance of $a=0.142nm$. As seen in Fig. 1(a), the
line defect is adjacent to two different sublattice sites (A and B).
In Fig. 1(b) the line defect connects two same-type sublattice sites
(either A-A or B-B). For simplicity, we call the structure depicted
in Fig. 1(a) the configuration 1 and that in Fig. 1(b) the
configuration 2 in the following discussion. Furthermore, it is easy
to find the main difference between the two types that mirror
symmetry with respect to the line defect is broken in the
configuration 1, but is kept for the configuration 2.

In this section, results of numerical calculations for the band
spectrum, the characteristics of a line defect, the distribution of
wave functions in real space, and the conductance are reported.

\subsection{The band spectra for various configurations}
In Fig. 2, we plot the band spectra of the two configurations shown
in Fig. 1 with different parameters. Note that here $U_d$ is set
zero in all cases of Fig. 2. First, to easily capture the changes in
the band structure induced by a staggered potential, the band
spectrum with a line defect is firstly plotted with parameter
$\Delta=0$ in Fig. 2(a). It can be found that two additional states
(red lines) are included in contrast to the band structure of
pristine graphene.~\cite{Cresti} Note that due to the two additional
states the particle-hole symmetry is broken, which probably gives
rise to some intriguing properties in this
system.~\cite{Gunlycke,Bahamon,Okada,Kou} In Fig. 2(b-d), a bulk
band gap can be observed for two configurations when choosing a
nonzero staggered potential. Importantly, when $\Delta=0.3t$ for the
configuration 1, an extra state due to the line defect can be seen
in the gap and fills approximately half of the band gap. However,
this special state in the band gap due to the line defect disappears
completely when we turn to the band spectrum of the configuration 2
with $\Delta=-0.3t$ as shown in Fig. 2(c). Furthermore, this state
appears again and becomes gapless when $\Delta$ is tuned to the
value of $0.3t$ for the configuration 2 in Fig. 2(d)

\begin{figure}[!ht]
\includegraphics[width=1.0\columnwidth, viewport=424 45 950 530, clip]{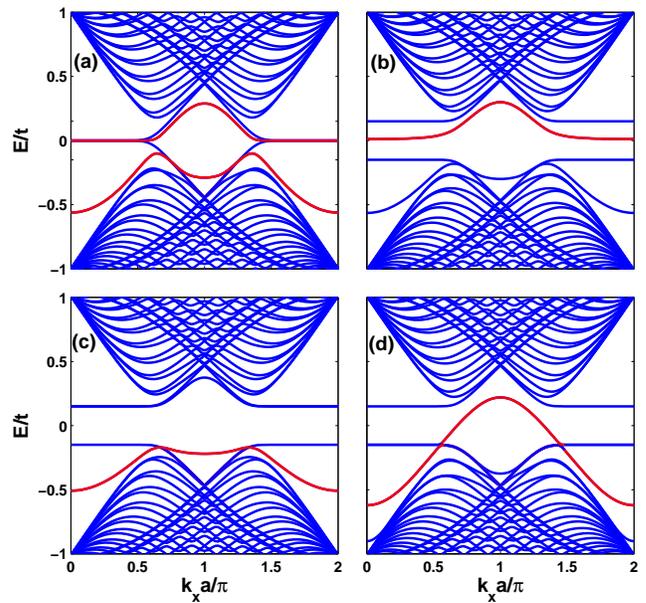}
\caption{ (Color online) Band spectrum in Fig. 2. (a) is plotted
with $\Delta=0$ for the configuration in Fig. 1(a) or 1(b). (b)
corresponds band spectrum of the configuration 1 shown in Fig. 1(a)
at $\Delta=0.3t$; (c) and (d) are plotted with $\Delta=\mp0.3t$
respectively for the configuration 2 shown in Fig. 1(b). For all
curves, the onsite energy of carbon atoms in the line defect $U_d$
is  zero and the width of ribbon is chosen to be $M=20$.}
\end{figure}

\subsection{The properties of the gapless state}
Some questions from these new results naturally arise such as: What
characteristics does the special state due to the line defect shown
in Fig. 2(d) have, and are there some novel applications in graphene
electronics related to these different band spectra? These are the
questions of interest in this paper. For simplicity, we will focus
mainly on the band spectrum in Fig. 2(d) with $\Delta=0.3t$ because
the gapless state has, in general, much more fascinating physics and
all band spectra should have the same origin. In fact, the following
discussions can also apply to other cases equally.

\begin{figure}[!ht]
\includegraphics[width=1.0\columnwidth, viewport=368 138 870 411, clip]{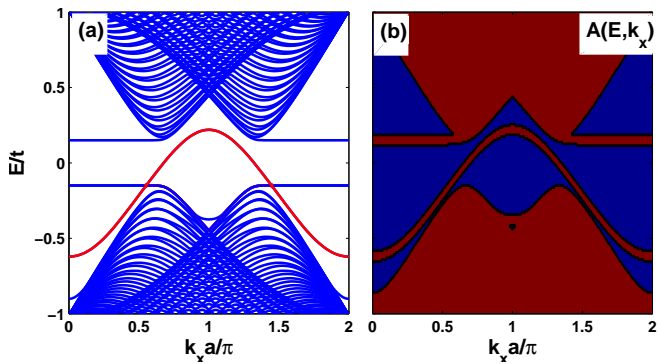}
\caption{ (Color online) Band spectrum (a) is plotted with
$\Delta=0.3t$ and $M=40$; (b) exhibits the local density of states
$A(E,k_x)$ vs $E$ and $k_x$ for an infinitely wide graphene sheet with
a line defect, namely $M=\infty$. Note that, for both (a) and (b), the onsite energy of carbon atoms in the
line defect $U_d$ is set zero and only the configuration 2 in Fig.
1(b) is considered.}
\end{figure}

Sometimes, the band spectrum for an infinitely long ribbon is
dependent on ribbon's width. For example, graphene ribbon with
armchair edges exhibits the characteristics of a metal or an
insulator depending on the ribbons' width. This motivates us to
study the dependence of the band spectrum on the width of ribbon. In
contrast to the width of ribbon with $M=20$ in Fig. 2(d), the
gapless state still exists and no noticeable changes can be observed
in the whole band spectrum when setting $M=40$ in Fig. 3(a). In
fact, regardless of what the width of the ribbon is set, the band
spectrum looks similar to that in Fig. 3(a).

Furthermore, to provide
a further evidence that the gapless state does not disappear and
the system at this case always keeps metallic, Fig. 3(b) shows the local density
of states (LDOS) nearby the line-defect carbon atoms for an infinitely wide graphene with a line defect (namely, $M=\infty$).
Note that LDOS nearby the line defect can be
detected experimentally by a scanning tunneling
microscope (STM) while in this paper LDOS is obtained theoretically
through calculating the Green's function of the line defect.~\cite{Jiang} In Fig. 3(b), we simply represent LDOS using two different colors,
namely the dark red color means a nonzero density of states
while the blue color represents a zero value.
As shown in Fig. 3(b), besides the disappearance of
edge states, LDOS nearby the line defect
exhibits nearly the same characteristics as the band spectrum in Fig.
2(d) or Fig. 3(a). The results of LDOS also illustrates that the transport
properties of this system are independent of the width of ribbon and
should be only related to the line defect.

\begin{figure}[!ht]
\includegraphics[width=1.0\columnwidth, viewport=372 136 870 413, clip]{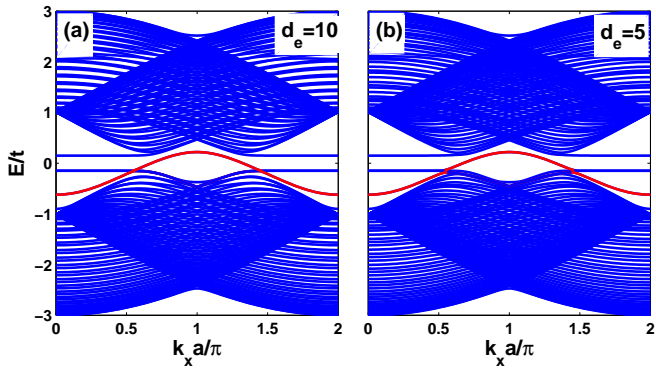}
\caption{ (Color online) Band spectra are plotted with two different
values of $d_e$. Here, $d_e$ denotes the distance between the line
defect and the top edge of the ribbon. (a) and (b) correspond to
the cases of $d_e=10$ and $d_e=5$, respectively. In all curves, only the
configuration 2 in Fig. 1(b) is considered. Other relevant
parameters are $\Delta=0.3t$, $M=40$ and $U_d=0$.}
\end{figure}

All the above calculations are made on the geometry that the line
defect lies in the middle of graphene ribbon. Whether the
defect-to-edge distance would affect the band spectrum is worth
discussing. Motivated by this idea, we next describe a study of the
influence of different defect-to-edge distances on the band spectrum
and results are shown in Fig. 4. Here, the distance from the line
defect to the top edge of the ribbon is represented as $d_e$, such
as $d_e=20$ in Fig. 3(a), which means the line defect is exactly in
the middle of the ribbon. When shifting the line defect toward the top
edge of ribbon with $d_e=10$, no obvious changes could be seen in the band spectrum
in Fig. 4(a). A similar band spectrum shown in Fig. 4(b) is also
observed when the line defect is moved much closer to the top edge
of ribbon with $d_e=5$. Further calculations with other different
$d_e$ show that the band spectrum keeps no noticeable changes if the line
defect is not very close to the edge of the ribbon. Therefore, where
the line defect lies in the ribbon has almost no influence on the band
spectrum, especially the gapless state. Meanwhile, this conclusion can be also applicable to all cases in Fig. 2.

\begin{figure}[!ht]
\includegraphics[width=1.0\columnwidth, viewport=370 130 882 410, clip]{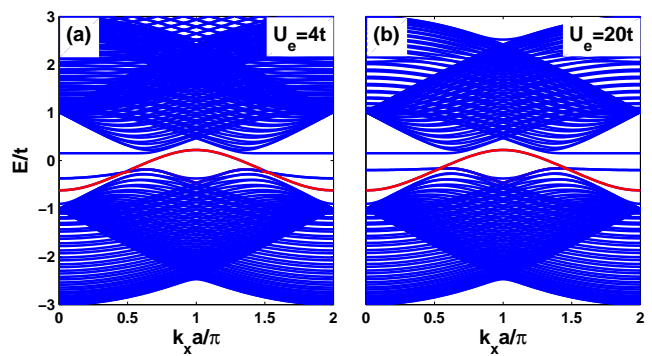}
\caption{ (Color online) Band spectra are plotted for the
configuration 2 with two different values of $U_e$, (a)$4t$ and
(b)$20t$. Here, $U_e$ denotes an additional onsite energy for the
top (down) edge of ribbon with width $M_e=10$. In all curves, only
the configuration 2 is considered and other parameters are set
$\Delta=0.3t$, $M=40$, and $U_d=0$.}
\end{figure}

From the discussions above, it is clear that band spectra in Fig. 2
are independent of ribbon's width and the location of the line defect.
To further verify that the gapless state in Fig. 2(d) is not an edge
state and not affected by the edge, we consider an additional onsite
energy $U_e$ on both edges of the ribbon. Note that the additional onsite
energy can be induced due to the surrounding environment or other
factors, such as chemical passivation and roughness at the
edges.~\cite{Cresti} Here we use the parameter $M_e$ to denote the width of the top (down) edge
where the additional onsite energy $U_e$ is induced. In these calculations we set the width of
ribbon $M=40$ and a symmetrical configuration is considered.

Fig. 5 shows expanded views of the band spectrum centered on the
energy $E=0$. As expected, no obvious changes in the band
spectrum, especially the gapless state, can be seen in Fig. 5(a)
with $U_e=4t$ or in Fig. 5(b) with $U_e=20t$. Thus, the gapless
state in Fig. 2(d) should not be an edge state. Besides,
it can be concluded from the results in Fig. 5 that the gapless
state keeps robust against the variation of edges.

\begin{figure}[!ht]
\includegraphics[width=1.0\columnwidth, viewport=300 240 798 510, clip]{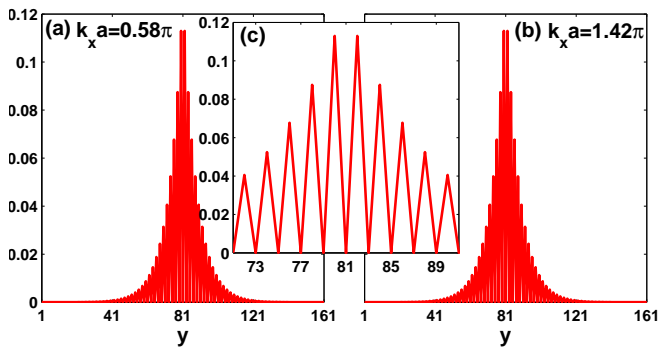}
\caption{ (Color online) Distribution of the wave function in real
space for two different cases, $k_x=0.58\pi$ and $k_x=1.42\pi$ is
considered in (a) and (b) respectively.
 Distribution of the wave functions are plotted only for the configuration 2 under condition:
$\Delta=0.3t$, $E=-0.12t$, $U_d=0$, and $M=40$. (c) is a magnification of (a).}
\end{figure}

As is known, an effective method to investigate the gapless state in
Fig. 2(d) is to examine the distribution of wave functions in real
space. In Fig. 6, we investigate the distribution of wave functions for two
different momentums. Here, the energy is set $E=-0.12t$, which
corresponds an energy level inside band gap and intersects with
gapless state at two momentums of $k_x=0.58\pi$ and $k_x=1.42\pi$.
The horizontal axis in Fig. 6 denotes carbon atom sequence across the ribbon width along the $y$ direction.
Note that the top lattice site of the ribbon corresponds to $y=1$, the bottom lattice site is presented by $y=161$,
and additionally the lattice site of the line defect is located at $y=81$.

In Fig. 6(a), when $k_x$ is chosen to $0.58\pi$, it can be observed
that the wave function mainly localizes around the center of ribbon
where the line defect is located. Furthermore, the completely same
distribution of the wave functions can be found for the case with
the momentum $k_x=1.42\pi$ in Fig. 6(b). To seen the wave
function's distribution more clearly, Fig. 6(c) shows a magnification of that in
6(a). Most important is that in Fig. 6(c) the wave function
is zero at the 81th site, which exactly corresponds to the site of
the line defect. That is to say, the neighbor sites to the line
defect mainly contribute to the gapless state inside gap. It is
shown by further numerical calculations that the wave function is
located mainly around the center of ribbon when the energy is near
to the top of valence band (namely, $E=-|\Delta|/2$). However, a
complex picture is obtained for the energy far from the top of the
valence band.

Based on discussions above, the
gapless state in Fig. 2(d) is stable and robust when changing the
width of ribbon, the onsite energy of ribbon's edges, and the
position of line defect within the ribbon.~\cite{Note1} Consequently, using the
line defect in graphene shown in Fig. 1, it may be possible to
construct a one-dimensional quantum channel to connect various
quantum devices. The technology for fabricating and controlling
graphene based electronic device is becoming excellent with the boom
of graphene research.~\cite{Neto0} In experiment, it would be
feasible to produce a quantum channel in graphene by using a line
defect method. Thus, line defects can have a significant
contribution in the application of graphene based electrical
systems.

\subsection{The variation of the band spectra with changing the onsite energy of line defect, $U_d$}
A fundamental question arises if a quantum channel is to be realized
using the line defect. That is, is there an easy way
to turn on and off  the quantum channel through some simple methods,
e.g. optical, electronic, and other ways? Motivated by this
idea, we study the behavior of the band spectrum when the
onsite energy of the line defect in graphene is tuned. In this
example, the onsite energy of the line defect could be tuned continuously by a gate
voltage embedded below the line defect. Here, other parameters are
set $\Delta=0.3t$ and $M=40$.  First, when the onsite
energy of line defect is set to be a very large negative value, e.g.
$U_d=-100t$, the graphene ribbon can be simply regarded as two
separated ribbons divided by the line defect. As seen in Fig. 7(a),
the band spectrum has similar characteristics as that of an
infinitely long graphene ribbon without a line defect. When changing
the onsite energy $U_d$ to the value of $-1t$, there is no big
change in the band spectrum. However, we can see that a state begins
to shift upward from the valence band when carefully examining the
top of valence band in Fig. 7(b).

\begin{figure}[!ht]
\includegraphics[width=1.0\columnwidth, viewport=285 60 730 536, clip]{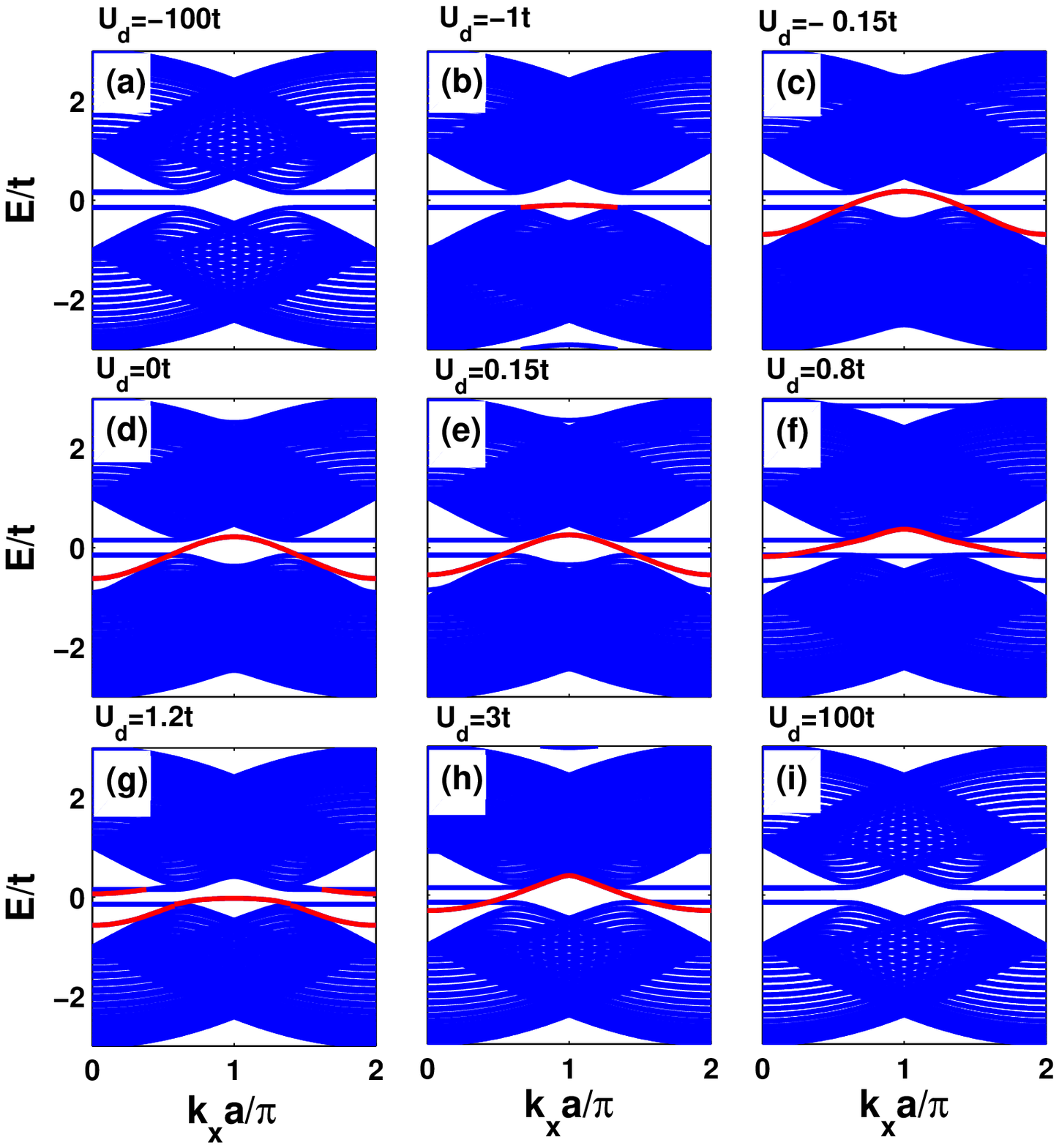}
\caption{ (Color online) Band spectra (a-i) are plotted with
changing onsite energy $U_d$ of the line defect. Only the
configuration 2 is considered. The parameters are chosen with
$\Delta=0.3t$ and $M=40$.}
\end{figure}

When tuning the onsite energy of the line defect to $U_d=-0.15t$ as
shown in Fig. 7(c), a clear gapless state can be observed
inside the band gap. Changing $U_d$ from $-0.15t$ to $0.15t$ as
shown in Figs. 7(c-e), the gapless state always exists and no
obvious change can be observed in the band spectrum. It is
interesting to note that the band gap due to staggered potential
is precisely $\Delta=0.3t$. By further increasing $U_d$, the state
inside the gap is shifted upward, but still remains gapless as in
Fig. 7(f). 

However, the band gap is half-filled by the state of line
defect once the onsite energy of line defect rises to the value of
$U_d=1.2t$ in Fig. 7(g). In addition, we observe that a new
state starts to emerge from the valence band in Fig. 7(g). Finally,
this new state of the line defect takes the place of the state
before and the bulk gap is occupied by a gapless state again in Fig. 7(h). With a
further growth of onsite energy $U_d$, this new state is shifted
upward and merges into the conduction band when the value of $U_d$
is large enough (approximately at $U_d\approx 50t$). The band
spectrum with a very large value of $U_e=100t$ is shown in Fig 7(i)
and appears very similar to Fig. 7(a), which exhibits a similar band
spectrum as that of a graphene ribbon without a line defect. Note that, 
here the cases of $U_d=\pm100t$ are studied to present what will happen 
for a very big value of $U_d$ in theory and the similar cases 
are also studied theoretically in Reference 48. In fact, we 
just need a small range of $U_d$, no more than $1t$, to tune the state of the
line defect from the band gap to the bulk.

\begin{figure}[!ht]
\includegraphics[width=1.0\columnwidth, viewport=285 60 730 536, clip]{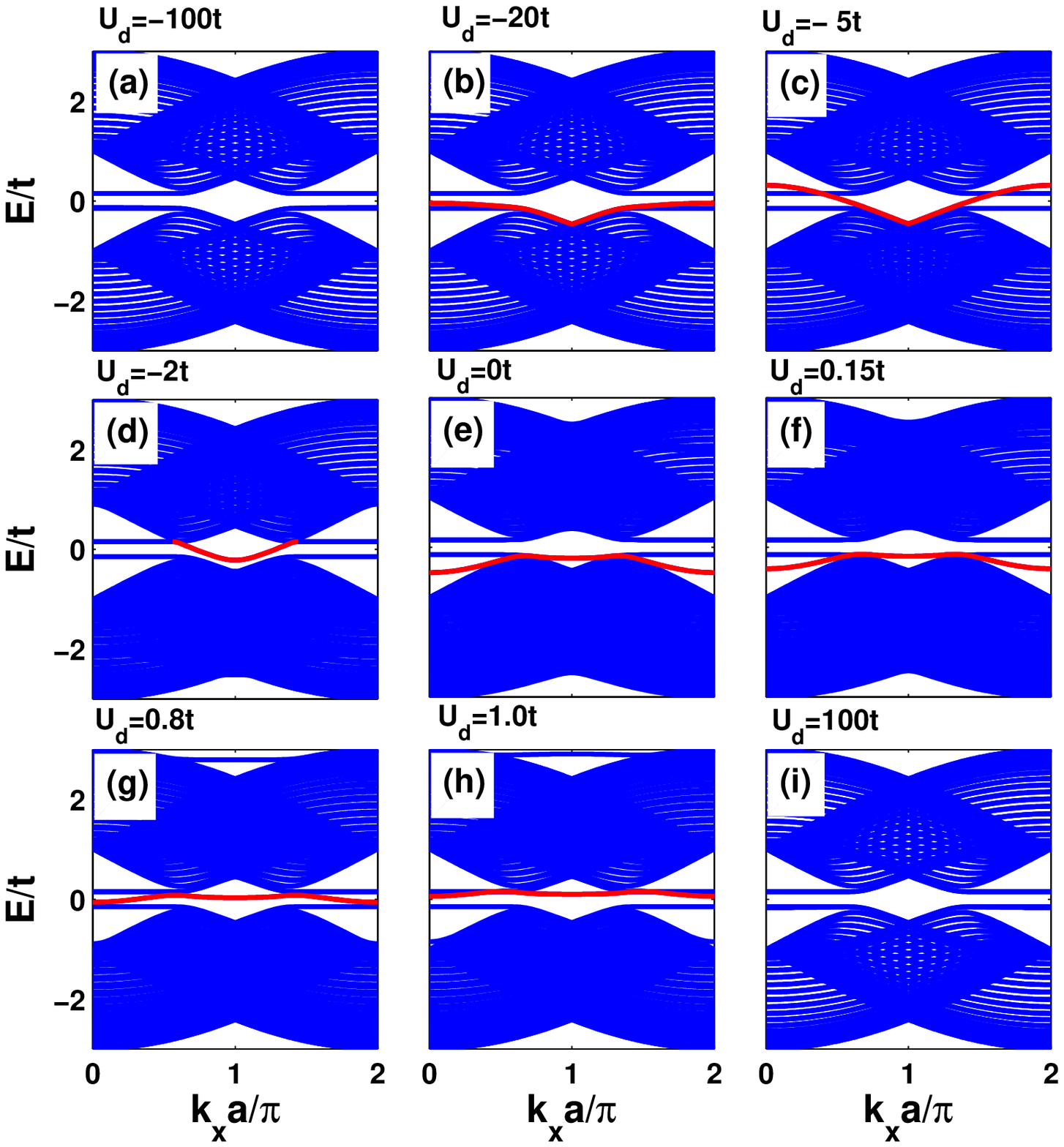}
\caption{ (Color online) Band spectra (a-i) are plotted with
changing onsite energy $U_d$ of the line defect. Only the
configuration 2 is considered. The parameters are chosen with
$\Delta=-0.3t$ and $M=40$.}
\end{figure}

Furthermore, when the sublattice potential is changed from $\Delta=0.3t$ (Fig. 7) to $\Delta=-0.3t$ (Fig. 8),
similar behaviors of band spectra can be found in Fig. 8. Note that, a excellent gapless state can be observed
at $U_d=-5t$, Fig. 8(c); two states of line defect are shifted from the valence band to the conduction band when
increasing the onsite energy of line defect $U_d$ gradually.
For simplicity,  behaviors of band spectra for the configuration 1 are not shown here. Note
that, for the configuration 1 with parameters $\Delta=0.3t$ and $M=40$, only one state of line defect
is shifted from the valence band to the conduction band and it never become gapless when increasing the onsite energy of line defect.
To sum up, the state inside the gap could be
tuned by the onsite energy of the line defect, which can be realized
by a gate voltage embedded below the line defect.

\subsection{The conductance at various cases}
By the Landauer-B\"{u}ttiker formula, the linear conductance of a mesoscopic system at zero
temperature and low bias voltage can be represented
as:~\cite{Song,Pareek,SDatta}
\begin{eqnarray}
G_{LR} &
=&\frac{e^2}{h}T=\frac{e^2}{h}Tr{[Re(\Gamma_LG^r\Gamma_RG^a)]},
\end{eqnarray}
where $T=Tr{[Re(\Gamma_LG^r\Gamma_RG^a)]}$ is the transmission
coefficient from the left lead (source) to the right lead (drain),
$G^{r/a}$ is the retarded/advanced Green's function and
$\Gamma_{L/R}=i(\Sigma^r_{L/R}-\Sigma^a_{L/R})$ with
the retarded/advanced self-energy $\Sigma^{r/a}_{L/R}$. For simplicity, The
size of the central region is denoted by integers M and N, which
represent the width and length, respectively. For example, in Fig. 1,
the width $L_y \approx M\times 3a$ with $M=4$, and the length
$L_x=N\times 2\sqrt{3}a$ with $N=5$.

\begin{figure}
\includegraphics[width=1.0\columnwidth, viewport=62 62 764 591, clip]{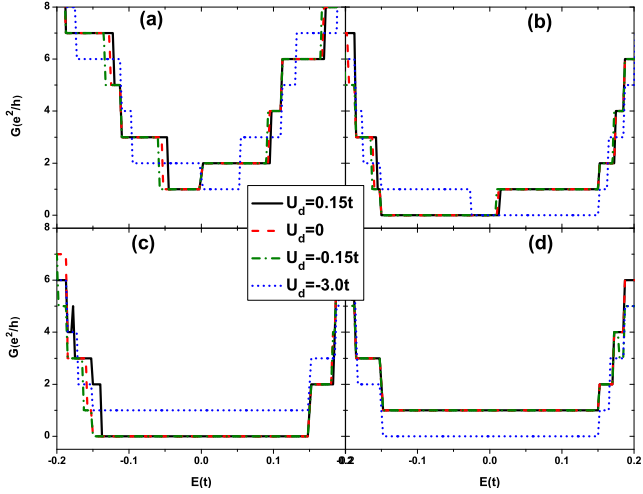}
\caption{ (Color online) Linear conductance vs Fermi level is
plotted for various onsite energies for a line defect. (a)
corresponds to the first or second configuration with $\Delta=0$;
(b) is plotted for the first configuration with $\Delta=0.3t$. (c)
and (d) correspond to the configuration 2 with $\Delta=-0.3t$
and $\Delta=0.3t$ respectively. }
\end{figure}

In Fig. 9, we plot the conductance for two configurations with
different staggered potentials and onsite energies for the line
defect. There are four main conclusions that can be made from the
results shown in Fig. 9. (1) For the configuration 1 or 2 with
$\Delta=0$ in Fig. 9(a), an asymmetrical conductance plateau is
observed near the Dirac point, which indicates that the
particle-hole symmetry is broken due to the existence of line
defect. Because there is no band gap at this case, the conductance
would change but always keeps no less than $G_0=e^2/h$ for the
different onsite energies of the line defect, $U_d$. (2) For the
configuration 1 with $\Delta=0.3t$, the conductance in Fig. 9(b)
jumps from zero to a $G_0$ quantum plateau for $U_d=0, \pm0.15t$
while for $U_d=-3t$ it falls from a $G_0$ quantum plateau to zero
when increasing the Fermi energy from $-0.15t$ to $0.15t$. (3) In all plots shown
in Fig. 9, the conductance curves keep almost unchanged for $U_d=0,
\pm0.15t$. However, for $U_d=-3t$, the conductance always shows
great differences from that at the other cases with $|U_d|\leq\Delta/2$. This
is because, in this case, the line-defect state is greatly affected
by such large onsite energy of the line defect and thus far away
from the original position. (4) Specially, it is observed in
Figs. (c) and (d) that the conductance exhibits either a quantum
plateau or zero for the energy inside the band gap. For example,
in Fig. 4(d), the $G_0$ quantum plateau
has almost no variation when $|U_d|\leq 0.15t$. However, the
conductance falls to zero for a large onsite energy of the line
defect, $U_d=-3t$, because here the line-defect state is far from
the gap and completely merges into the bulk band.

Note that all of
the observations about the conductance in Fig. 9 are consistent with the
characteristics of the band spectra discussed above in Fig. 7 and Fig. 8. In summary,
the graphene with a line defect can exhibit an insulator or a metal
with a quantum conductance $e^2/h$. Therefore, a quantum channel can
be constructed using the line defect and importantly it can be
turned on or off easily by changing the onsite energy of the line
defect.

\begin{figure}
\includegraphics[width=1.0\columnwidth, viewport=72 300 764 591, clip]{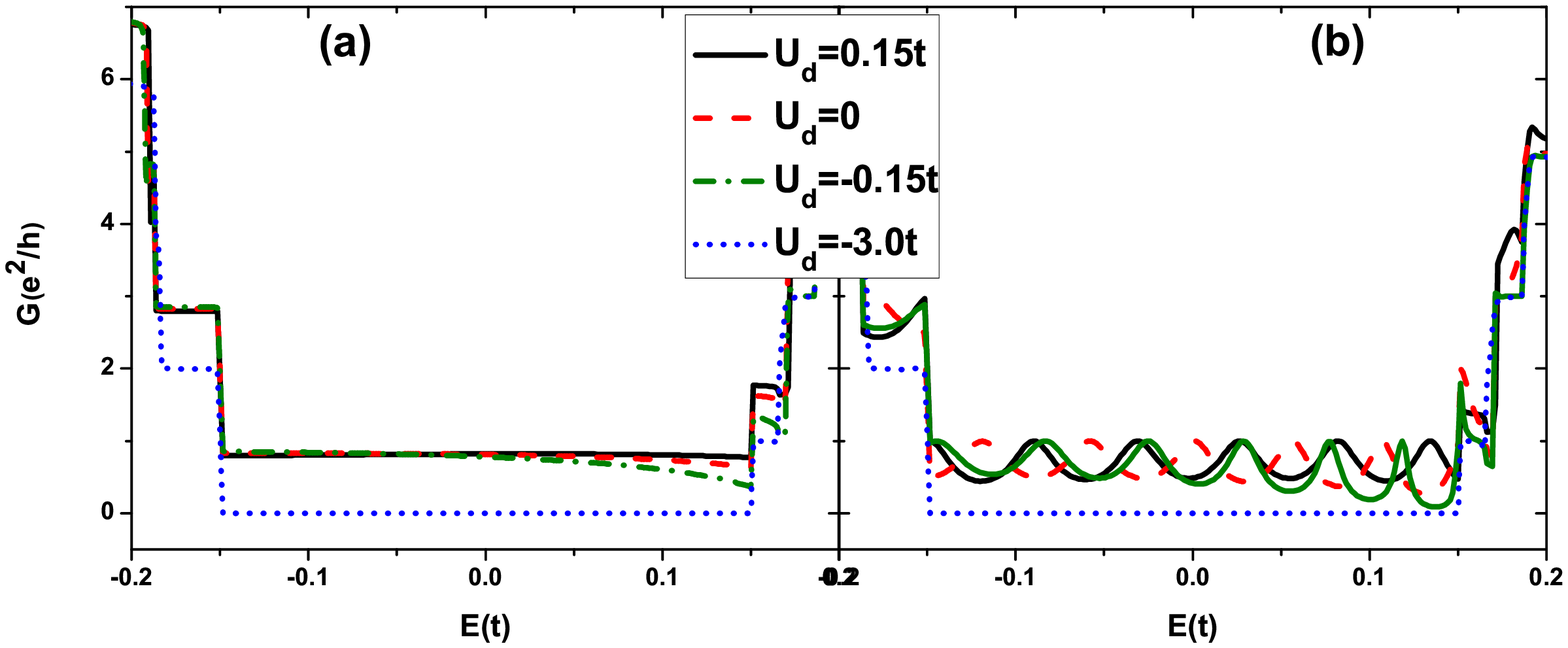}
\caption{(Color online) The influence of vacancy on the conductance for a line defect.
(a) is plotted for single vacancy; (b) is for two vacancies, the distance of which is set $N_c=20$.
All conductances are calculated for the configuration 2 with $\Delta=0.3t$, $M=40$, and $N=21$.}
\end{figure}

In general, vacancies or point defects can affect the transport of a line
defect greatly.~\cite{Bahamon} Thus, we consider the influence of vacancy on
the transport properties of a line defect and the results are shown
in Fig. 10. Here, we assume that a vacancy is generated due to two-atom missing from the line defect, which lie at the sites in the
same supercell (such as the two red sites in the red rectangle in
Fig.1). Meanwhile, in order to minimize the total energy, the carbon
atom's electrons, which lie at the up and down carbon atoms sites
nearby the missing atoms, can now hop directly with
strength $t$. As shown in Fig. 10(a), although the conductance has a
dip to zero for Fermi energies near the bottom of the conductance band,
the quantum conductance plateau is robust for Fermi energies near
the top of the valence band. This conclusion is general for all
curves when $|U_d|\leq\Delta/2$. Additionally, when considering two
vacancies in a line defect, a completely different picture is presented in Fig. 10(b).
Here, the size of the central region is chosen to be $M=40$ and $N=21$,
and two vacancies of the line defect lie at both ends of  the line defect in the central region,
(the distance of two vacancies is therefore denoted by $L_c=N_c\times 2\sqrt{3}a$ with $N_c=20$).
As shown in Fig. 10(b), the quantum conductance plateau disappears and is replaced
by the conductance oscillations. That is because a standing
wave is formed due to reflections by the two vacancies. Therefore,
the oscillations shown in this energy range are simply the Fabry-Perot
interference effect due to the broken translational symmetry
introduced by the vacancies in the line defect.~\cite{Bahamon}

\section{CONCLUSIONS}
With considering a broken bulk inversion symmetry due to a staggered
potential, we study the graphene with a line defect and find that
distinctive band spectra could be obtained for different
configurations which can be attributed to different distributions of
a staggered potential. In particular, a gapless state exists for a
given configuration which holds a mirror symmetry with respect to the
line defect. Further calculations show that this special band
structure is independent of the width of ribbon, the onsite energy
of ribbon's edges, and the location of line defect within the
ribbon. Through analyzing the distribution of wave function of the
gapless state, it is found that this gapless state should be attributed
to the line defect. In addition, the gapless state is robust
against a weak perturbations of the onsite energy of the line
defect, e.g., keeps always gapless when $|U_d|\leq 0.15t$ [shown in Fig. 9].

These exotic characteristics of line defects illustrate that a
quantum channel could be fabricated using the line defect. In
particular, the turning on or off of this quantum channel can be
realized by a gate voltage embedded below the line defect. The line
defect could then be used to connect different quantum devices and
serve as a quantum channel to transport quantum signals between
devices. All the results in this study support the idea that the
line defect is a useful structure for the implementation of a
quantum channel in the field of microelectronics.

\section*{ACKNOWLEDGMENTS}
We are grateful to Tao Qin and Yanyang Zhang for their helpful
discussions. Juntao Song is supported by NSFC under Grant No.
11047131, No. 11147172, RFDPHE-China under Grant No. 20101303120005, and SFHP under Grant No.A2012205069. Hua Jiang
is supported by China Post-doctroal Science Foundation under Grant
No. 20100480147 and No. 201104030. Qing-feng Sun and X.C. Xie are supported by NSFC under
Grant No. 11074174, No. 10974236, and China-973 program.

\newpage

\end{document}